# Design of a plasmonic-organic hybrid slot waveguide integrated with a bowtie-antenna for terahertz wave detection

Xingyu Zhang*[a,c], Chi-Jui Chung[a], Harish Subbaraman[b], Zeyu Pan[a], Chin-Ta Chen[a], and Ray T. Chen*[a,b]

[a] University of Texas at Austin, 10100 Burnet Rd, MER 160, Austin, TX 78758, USA;
[b] Omega Optics, Inc., 8500 Shoal Creek Blvd, Austin, TX 78757, USA;
[c] Now with Acacia Communications, Inc., Hazlet, New Jersey 07730, USA.

## ABSTRACT

Electromagnetic (EM) wave detection over a large spectrum has recently attracted significant amount of attention. Traditional electronic EM wave sensors use large metallic probes which distort the field to be measured and also have strict limitations on the detectable RF bandwidth. To address these problems, integrated photonic EM wave sensors have been developed to provide high sensitivity and broad bandwidth. Previously we demonstrated a compact, broadband, and sensitive integrated photonic EM wave sensor, consisting of an organic electro-optic (EO) polymer refilled silicon slot photonic crystal waveguide (PCW) modulator integrated with a gold bowtie antenna, to detect the X band of the electromagnetic spectrum. However, due to the relative large RC constant of the silicon PCW, such EM wave sensors can only work up to tens of GHz. In this work, we present a detailed design and discussion of a new generation of EM wave sensors based on EO polymer refilled plasmonic slot waveguides in conjunction with bowtie antennas to cover a wider electromagnetic spectrum from 1 GHz up to 10THz, including the range of microwave, millimeter wave and even terahertz waves. This antenna-coupled plasmonic-organic hybrid (POH) structure is designed to provide an ultra-small RC constant, a large overlap between plasmonic mode and RF field, and strong electric field enhancement, as well as negligible field perturbation. A taper is designed to bridge silicon strip waveguide to plasmonic slot waveguide. Simulation results show that our device can have an EM wave sensing ability up to 10 THz. To the best of our knowledge, this is the first POH device for photonic terahertz wave detection.

The detection and measurement of electromagnetic fields over a large spectrum have attracted significant amounts of attention in recent years. Traditional electronic electromagnetic field sensors [1, 2] use large active conductive probes which perturb the field to be measured. This also makes the devices bulky with strict limitation on RF bandwidth. In order to address these problems, integrated photonic electromagnetic field sensors [3-6] have been developed, in which an optical signal is modulated by an electric field collected by a miniaturized antenna. In our recent work [7], we have designed, fabricated and experimentally demonstrated a compact, broadband and highly sensitive integrated photonic electromagnetic field sensor to cover the X band of the electromagnetic spectrum (8-12GHz). This photonic electromagnetic field sensor consists of an electro-optic (EO) polymer refilled slot photonic crystal waveguide (PCW) optical modulator coupled with a gold bowtie antenna, as shown in Figs.1 (a) and (b). The frequency was targeted around 10GHz which is the central frequency of X band. However, due to the intrinsic limitation of silicon photonic crystal waveguide based structure with relative large RC time constants, such RF sensors can only go up to tens of GHz. To overcome this limitation, our group first proposed the idea of using EO polymer refilled plasmonic slot waveguide in conjunction with a specially shaped bowtie antenna to cover wide electromagnetic spectrum including the range of microwave, millimeter wave and even terahertz waves [8]. This work was then followed by several good experimental demonstrations [9, 10]. In this paper, we present detailed design, simulation and analysis of this organic-plasmonic hybrid (POH) slot waveguide integrated with the bowtie antenna.

One feasible design is shown in Figs.1 (c) and (d), in which the plasmonic slot waveguide sensor infiltrated with EO polymer can have a detection bandwidth of up to THz frequency range. The bowtie antenna has a conventional bowtie shape with extension bars attached to its apex points [11, 12], and it serves as a receiving antenna, a driving electrode, and a poling electrode. The special feature of this bowtie antenna include simple planar structure, stable broadband performance, and strong local near-field enhancement [13]. Under RF illuminations, this bowtie antenna concentrates the energy and provides an extended near-field area with a uniformly enhanced electric field. The field enhancement by the bowtie antenna is about ~500 times based on numerical simulations. The bowtie antenna in the device is experimentally demonstrated to have broadband characteristics with a central resonance frequency of tens of GHz, as well as a large half-power beam width of 90 degree which enables the detection of electromagnetic waves from a large range of incident angles. And also, the resonant frequency of a bowtie antenna can be tuned by appropriately modifying and scaling the bowtie geometry, such as arm length, flare angle, and feed gap width [11]. This enables the bowtie antenna to have various applications over a wide frequency range to 10 THz, including sea surface perturbation detection through water molecule emission at 183 GHz [14], extreme-ultraviolet light generation [15], optical sensing and energy harvesting [16, 17].

*xzhang@utexas.edu
*chenrt@austin.utexas.edu

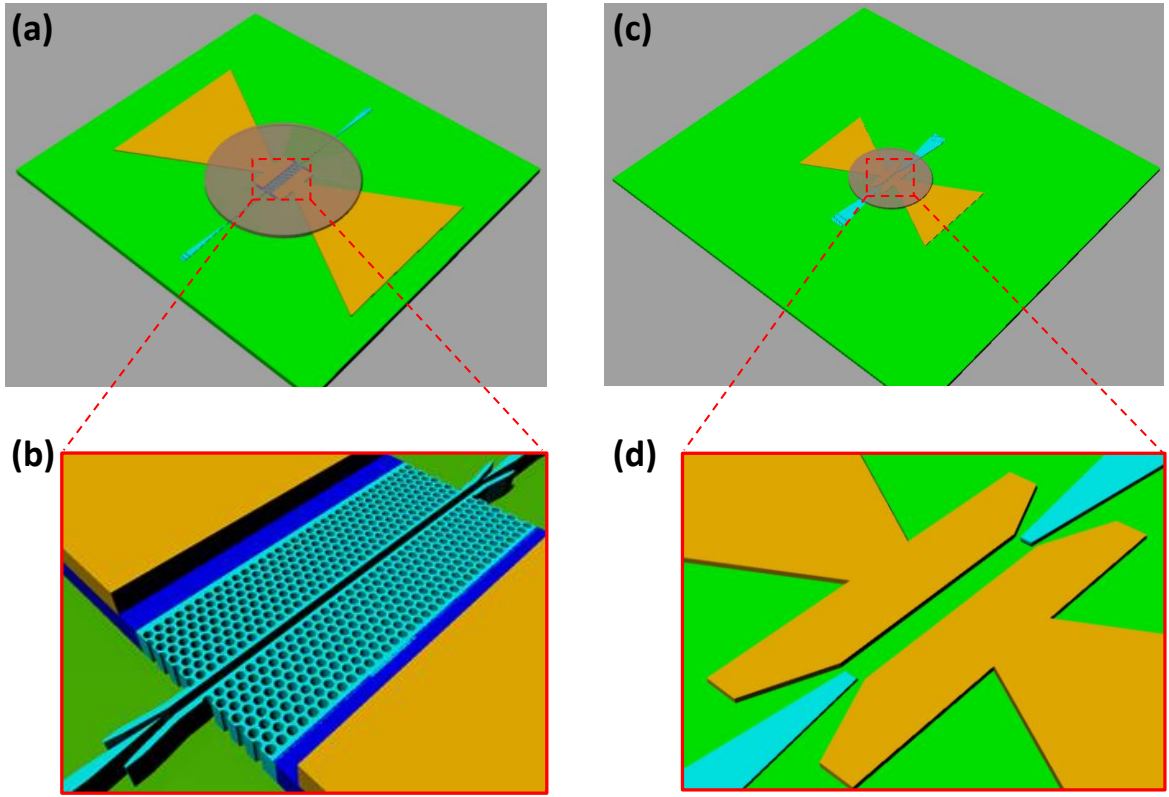

FIG. 1. (a) A schematic view of our electromagnetic field sensor consisting of an EO polymer refilled silicon slot PCW phase modulator and a bowtie antenna. (b) Magnified image of slot PCW. (c) A schematic view of the electromagnetic field sensor consisting of an EO polymer refilled plasmonic slot waveguide phase modulator and a bowtie antenna. (d) Magnified image of plasmonic slot waveguide. Note: the EO polymer layer covered on top of the device is not shown in (b) and (d) for better visualization.

The silicon slot PCW phase modulator we demonstrated previously was embedded in the feed gap of a good bowtie antenna [Figs. 1 (a) and (b)]. The doped silicon slot can further squeeze the electric field into the slot and further improve the local electric field enhancement by another ~20 times, leading to a total enhancement factor of ~10,000 (500×20). The silicon slot and PCW holes are filled with the EO polymer (SEO125 from Soluxra, LLC.) which has a large EO coefficient ($r_{33}$ of ~100pm/V at 1550nm), ultrafast response time (~several fs), low optical loss, as well as excellent thermal- and photochemical-stability [6]. The refractive index of the EO polymer can be changed by applying an electric field via the Pockels effect. Organic EO polymer has been used in many applications such as optical interconnects and communications [18, 19]. The hybrid integration of EO polymer with silicon photonics [20] combines the benefits of both platforms [21, 22]. The slow-light effect in the PCW can enhance the interaction of RF field and optical waves and thus increases the effective in-device EO coefficient of this modulator to be an experimentally demonstrated value of $r_{33}=1230$pm/V [23], which is 40 times larger than conventional Lithium Niobate (LiNbO$_3$) modulator ($r_{33}=30$pm/V) [24]. This promises a very high sensitivity of the device to the incident electric field. In addition, the silicon layer is selectively doped for high frequency operation [25], and modulation response up to 43GHz is measured, with a 3-dB bandwidth of 11GHz is experimentally demonstrated [26]. The sensor is experimentally demonstrated with a minimum detectable electromagnetic power density of 8.4mW/m$^2$ at 8.4GHz, corresponding to a minimum detectable electric field of 2.5V/m [7]. To the best of our knowledge, this was the first silicon-polymer hybrid device and also the first PCW device used for the photonic sensing of electromagnetic waves.

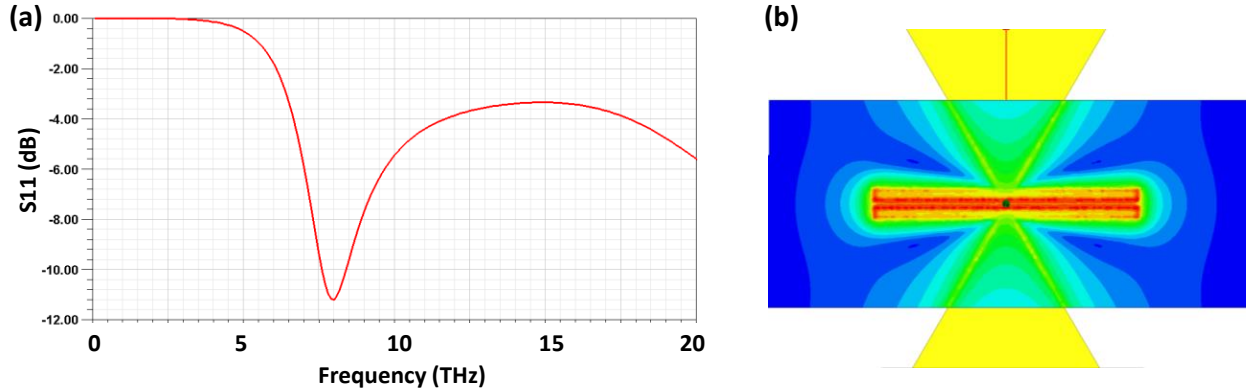

FIG. 2. (a) Simulated S11 parameter over a frequency range 1-20THz. (b) The middle region of the bowtie antenna structure, overlaid with simulated normalized electric field distribution at 8THz.

In order to make our sensor cover a large frequency regime up to 10 THz, the device is modified from a conventional waveguide to a gold plasmonic slot waveguide, as shown in Figs. 1 (c) and (d). First, the arm length of the bowtie antenna is scaled to provide a broad bandwidth with central resonance frequency in terahertz range [27]. Figure 2 (a) shows the S11 parameter of the bowtie antenna simulated by using ANSYS HFSS, with a resonance frequency around 8THz, and Fig.2 (b) shows the simulated normalized electric field distribution on bowtie antenna at 8THz with a strong near-field enhancement inside the feed gap.

Second, the silicon slot PCW in Ref. [7] is removed and then the gap between the two extended metal bars are narrowed down to 250nm to form a plasmonic slot waveguide. The slot between two metal bars has a narrow enough width to support a single mode surface plasmonic mode. In our design, the slot width is chosen to be 250nm and the gold thickness is chosen as 250nm, which is the same as the silicon top layer thickness. And also, the two metal pads also serve as poling electrode and modulation electrode. The metal slot is filled with EO polymer, and optical modulations can be achieved in this plasmonic slot waveguide as demonstrated recently in Ref. [28]. Figure 3 (a) shows the Lumerical simulation results of a surface plasmonic mode profile at the cross section of the EO polymer filled gold plasmonic slot waveguide, from which it can be seen that most of optical power is confined inside the slot. Figure 4 (b) shows the COMSOL simulation results of electric field profile at the same device cross section. The electric is also concentrated inside the metal slot. This leads to an almost 100% interaction factor with optical mode, which is beneficial for highly efficient EO modulation.

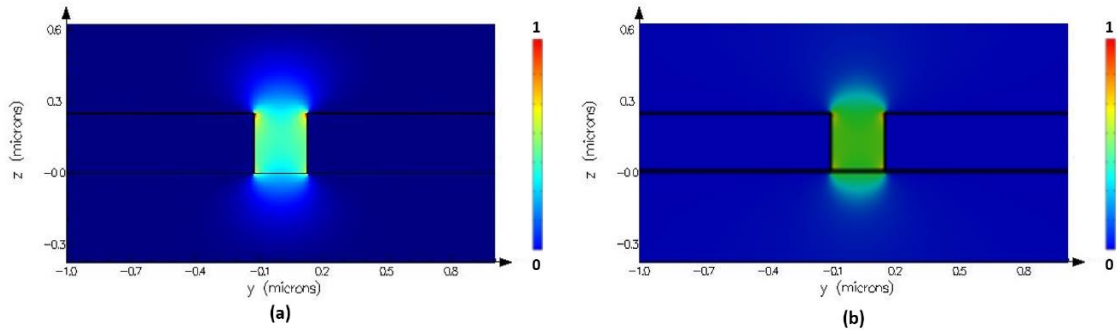

FIG. 3. (a) Simulated plasmonic slot waveguide mode profile. (b) Simulated RF electric field profile on the plasmonic slot waveguide.

The working principle of this terahertz wave sensor is based on the electric field enhancement by the bowtie antenna and the electro-optic modulation via EO polymer inside the plasmonic slot waveguide, similar to the X-band electromagnetic wave sensor. This plasmonic-organic hybrid (POH) integration [29] have recently become a hot topic, enabling some high-performance devices such as high-speed modulators [30] and even modulator array for optical interconnects [31]. From material point of view, EO polymer has ultrafast response time in femtosecond scale, so it has been used for many THz applications [32, 33]. For example, [34], EO polymer with operational frequency bandwidth up to 1.6 THz and large EO coefficient as high as 200pm/V was reported in Ref. [16]. Therefore, here we propose the use of a POH integrated device for photonic detection of terahertz waves.

There are several important benefits of this plasmonic-based terahertz wave sensor proposed herein. (1) Due to the removal of silicon PCW and the high conductivity of metals, the RC time delay is significantly reduced, enabling a very broad frequency bandwidth. The resonance of the bowtie antenna can be engineered to cover a frequency range from 1GHz to 10THz. Based on simulation results as shown in Fig. 4 (a), the normalized electric field dropped across the metal slot is almost constant from the frequency of 1 GHz to 10THz. As a comparison, for silicon PCW slot, the normalized electric field across the decreases as the frequency increases due to the finite conductivity of the silicon. Its 3-dB bandwidth is about 30GHz [25] which limits the detectable frequency band of the sensor. Figures 4 (b) and (c) shows the electric field potential distribution on the plasmonic slot waveguide and the silicon slot PCW at frequencies up to 10THz. It can be seen that the most of the modulation field is still efficiently dropped across the plasmonic slot where EO polymer is filled. (2) The electric field enhancement factor is increased when the bowtie feed gap is reduced, so the 250nm-wide metal slot can provide higher local electric field enhancement which improve the sensitivity of this sensor. A 2X improvement in sensitivity is expected compared to 320nm-wide silicon slot PCW. The 250nm metal gap can be further optimized, and the sensitivity can be further improved by reducing the metal slot width. (3) The large overlap between the modulation electric field and the surface plasmonic mode enables more efficient modulation, which can reduce the required interaction length. Although plasmonic waveguides have higher propagation loss per unit length compared to regular waveguides, the short device length is helpful in reducing the total propagation loss. (4) By properly designing the plasmonic slot waveguide and engineering its dispersion diagram, a large group index can be achieved in the surface plasmonic mode and this slow-light effect can be used to enhance the modulation efficiency and the detection sensitivity. (5) Some existing electronicbased THz detectors require low temperature to reduce electron noise, so it required a cooling process such as liquid Nitrogen flow; on the contrary, our photonic (plasmonic) detection approach has good noise immunity. Our device can also reduce the impact of perturbing fields (EMI and EMP effects), since it is based on an optical modulation technique.

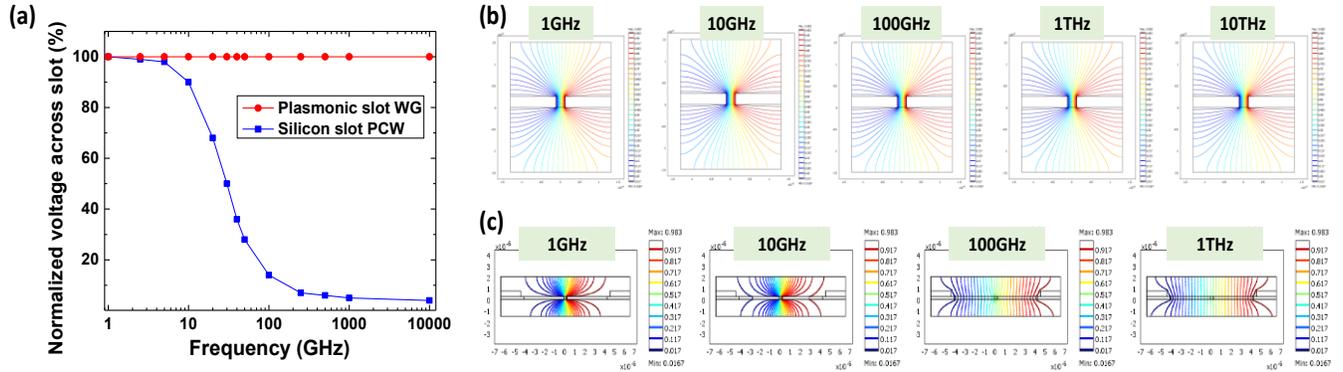

FIG. 4. (a) Simulated normalized voltage drop across the silicon slot PCW and the plasmonic slot waveguide. (b) Simulated electric potential distribution on the plasmonic slot waveguide, at 1GHz, 10GHz, 100GHz, and 1THz, respectively. (c) Simulated electric potential distribution on the silicon slot PCW, at 1GHz, 10GHz, 100GHz, respectively.

To analysis the optical loss of our device, the confinement factor of the optical mode of the plasmonic slot waveguide depending on the slot width is numerically calculated using Lumerical. Figure 5 (a) shows the confinement factor as a function of metal slot width. There is an optimized slot width (~60nm) for the maximum confinement factor of 0.75. Figure 5 (b) shows the dependence of group index on slot width. The smaller slot width leads to a higher group index, indicating a slow-down of energy velocity. The propagation loss of the plasmonic mode is also investigated in Fig. 5 (c). The optical loss is increased for smaller slot width. From electrical point of view, by reducing the slot width, the electrical field increase inversely. This gain in nonlinearity ultimately overcompensates for losses. Based on the simulated complex effective index, the propagation loss of the 250nm-wide plasmonic slot waveguide is below 0.4dB/μm. This propagation loss can be further reduced by the optimization of the plasmonic slot waveguide, the selection of metal with lower propagation loss (i.e. copper to replace gold), as well as improved fabrication quality provided by CMOS foundry. In Ref. [28], an interaction length of only 29 μm and a total insertion loss of 12dB are demonstrated. We expect the total insertion loss of our device can be controlled below 10dB.

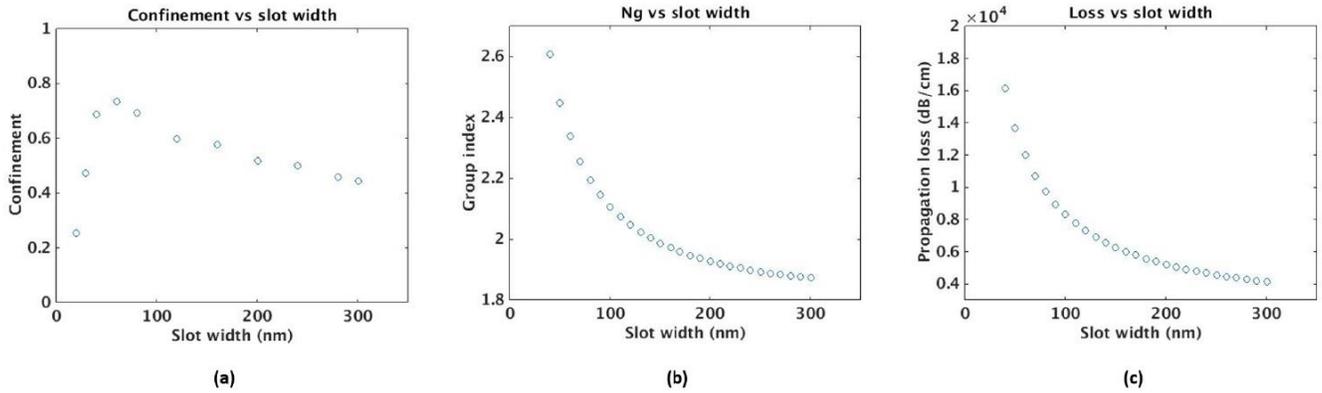

FIG. 5. (a) Simulated the confinement factor as a function of metal slot width. (b) Simulated group index as a function of slot width. (c) Simulated propagation loss of plasmonic mode as a function of metal slot width.

To bridge a silicon strip waveguide to the plasmonic slot waveguide as shown in Fig 1 (d), a hybrid silicon-gold taper is designed to efficiently couple light between conventional single mode and surface plasmonic mode [35], as shown in Figs. 6 (a)-(d). The silicon strip waveguide has a width of 450nm and thickness of 250nm. Considering the practical limitations on fabrication, the silicon taper width at the tip is assumed to be 60 nm. The plasmonic slot waveguide has a slot width of 250nm and thickness of 250nm. The EO polymer cladding has a thickness of 2 μm. The mode converter in Fig. 6 (a) has a length of 1.7μm. This strip-to-plasmonic slot waveguide mode converter is studied with three-dimensional (3-D) finite-difference time-domain (FDTD) simulations in Rsoft. Figure 6 (e) shows the field distribution ($E_x$) in the x-z plane at the wavelength of 1550nm, which shows the evolution of the optical mode from the silicon strip waveguide to the plasmonic slot waveguide via the hybrid silicon-gold taper. A very high conversion efficiency of 93.3% at 1550 nm is achieved. The detailed optimization of this hybrid silicon-gold taper can be found in Ref. [35].

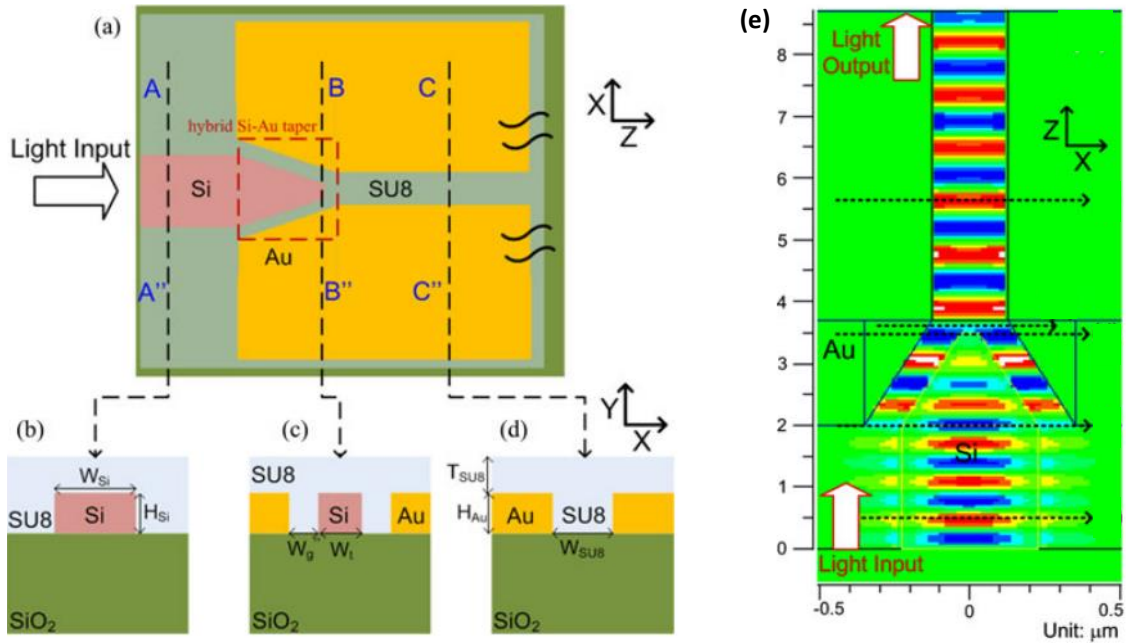

FIG. 6. Schematic diagrams of dielectric-to-plasmonic slot waveguide mode converter based on a hybrid silicon-gold taper. (a) Top view of the mode converter with a silicon strip waveguide at the input and a plasmonic metal slot waveguide at the output. Cross sections of (b) a silicon strip waveguide, (c) a hybrid silicon-gold taper, and (d) a plasmonic slot waveguide. (e) Top view of the field distribution (Ex) of mode converter at 1550 nm.

In conclusion, we present the detailed design and discussion of a terahertz wave sensor based on a plasmonic-organic hybrid (POH) slot waveguide modulator integrated with a bowtie antenna. Simulations show that the device provide a large overlap between plasmonic mode and RF field, an ultra-small RC constant, and strong electric-field enhancement, enabling an efficient and broadband optical modulation and thus high EM wave sensing ability up to 10 THz. A taper is designed to bridge silicon strip waveguide to plasmonic slot waveguide. To the best of our knowledge, this is the first detailed proposal of POH integrated terahertz wave sensor. The working frequency bandwidth of this sensor can be tuned by modifying the bowtie antenna geometry, in order to enable potential applications in other frequency ranges such as microwave detection, millimeter wave imaging, even light trapping for photovoltaics, and so on.

In the future work, the sensitivity can be doubled by integrating two of these devices in an Mach–Zehnder interferometer (MZI) and implementing a push-pull configuration. The device can also be made into a polarization-independent sensing structure by making the two bowties perpendicular to each other. A star shape sensor array with maximum surface area coverage can also be designed to increase the sensitivity with polarization independence. In addition, the sensor can be made on silicon-on-sapphire substrate to avoid the unwanted reflections from silicon handle and thus further improve the device sensitivity [36]. The sensor can also be made on flexible substrate and compatible with ink-jet printing technologies [37, 38]. Very recently, new EO polymers with utrahigh $r_{33}$ value of 273 pm/V and high refractive index of 2.12 at the wavelength of 1300 nm have been achieved from monolithic glass, which represents a record-high $n^3 r_{33}$ figure-of-merit of 2601 pm/V with good temporal stability at 80 ºC [39]. Benefiting from the remarkable progress in EO polymer material development, the sensor performance is expected to be further improved.